# Correctness is not enough

*Louise Pryor*

*www.louisepryor.com*

**ABSTRACT**

*The usual aim of spreadsheet audit is to verify correctness. There are two problems with this: first, it is often difficult to tell whether the spreadsheets in question are correct, and second, even* if *they are, they may still give the wrong results. These problems are explained in this paper, which presents the key criteria for judging a spreadsheet and discusses how those criteria can be achieved*

**1 INTRODUCTION**

Many people are, quite rightly, worried that their spreadsheets are producing the wrong numbers. Results summarised by Panko [2000] indicate that about 80% to 90% of spreadsheets contain significant errors. These results are hard to interpret, as the definitions of "significant error" vary; and the samples may be biased, either towards spreadsheets that are thought likely to contain errors, or towards those that are considered particularly useful to their users, and that might therefore have been subject to more intense scrutiny.

Looking at the percentage of individual cells with errors leads to roughly similar conclusions. Field audits (again from Panko) show cell error rates from 0.38% up to 2.5%; laboratory studies indicate rates varying from 1.1 % to 21 %. Again, these results are difficult to interpret: are they percentages of all cells, cells containing formulae, or unique formulae? (If a formula is copied down a row or column, it may count as many formula cells, but only one unique formula).

Table 1 lists a typical sample of the spreadsheets that 1 have reviewed recently: they have between 3,000 and 150,000 formula cells, and between 35 and 350 unique formulae. If a proportion *p* of unique formulae has errors, then a spreadsheet with *n* unique formulae has a probability of $1 - (1 - p)^n$ of containing at least one error. The table shows the results for p = 1%. If we ignore SS2, which relied mainly on macro execution for its calculation (thus introducing another set of problems), the range of error probabilities is 78% to 97%. These rates are consistent with the error rates from the field audits.

|     | Formula cells | Unique formulae | Error probability |
|-----|---------------|-----------------|-------------------|
| SS1 | 67,013        | 351             | 97%               |
| SS2 | 10,587        | 37              | 31%               |
| SS3 | 144,378       | 284             | 94%               |
| SS4 | 30,252        | 209             | 88%               |
| SS5 | 20,938        | 260             | 93%               |
| SS6 | 3,245         | 164             | 81%               |

**Table 1 Spreadsheet error rates from cell errors**

It was difficult to test the predictions in Table 1 by determining whether the spreadsheets did actually contain errors. In order for this to be possible the spreadsheets must be both specified and auditable. Most spreadsheets are neither, as discussed in the next section.

Moreover, as discussed in section 3, a spreadsheet that is technically correct may still produce the wrong numbers. Possible causes include poor usability, maintainability, and performance, characteristics that are often considered to be optional extras, affecting productivity but not really central to spreadsheet quality. We'll discuss why this attitude is misguided.

## 2 CORRECTNESS

Most of the spreadsheets that I review, and a high proportion of all spreadsheets, can be viewed as financial models. As such, there are two levels to their correctness: whether the correct model is being used, and whether the model is implemented correctly.

In addition, many spreadsheets use rates, factors and other data imported from elsewhere. This is especially frequent in life insurance, but is common in other industries too.

This section considers the issues of specification and auditability and how they affect the overall correctness of a spreadsheet. In most cases, it is not enough for a spreadsheet to be correct; it must be seen to be correct too.

### 2.1 Specification

It is impossible to tell whether a spreadsheet is doing the right thing, in other words whether it is correct, unless you know exactly what it is meant to be doing. You need a specification; you can then check whether the specification has been implemented properly, and whether the specification is itself correct. Both these checks are needed.

Very few, if any, spreadsheets are adequately specified. Generally, the most you can hope for in the way of a specification is one along the lines of "Calculate the rates according to the inputs," without a hint of what formulae should be used. Although a vague specification is enough to settle some issues, there are often a number of equally plausible ways of modelling the business in question. Without a detailed specification it is impossible for any reviewer to tell whether the method actually used is the one that was intended.

Perhaps the worst example of a missing specification that I have recently encountered was in an industry where capital allowances are extremely significant. One of the key issues in the review was whether they were being handled properly in the spreadsheet. The trouble was that nobody was prepared to say how they should be being handled, and it turned out that nobody had told the developer of the spreadsheet either. However, they had been given another spreadsheet and told to handle them in the same way. Unfortunately the spreadsheet supplied was totally undocumented; moreover, it emerged during discussions that it had been obtained under dubious circumstances from a competitor.

The lack of a specification is, almost certainly, a leading cause of errors; simply clarifying what the spreadsheet is meant to do can often assist in making sure that it actually does it. However, although a specification is necessary it is certainly not sufficient. As Butler pointed out, even in a domain such as indirect taxation, with generally well documented calculation rules, errors are frequent [Butler 2000].

| Financing Assumptions | | |
|---|---|---|
| Financing flag | 1 | (0=NPP, 1=financing |
| LAUTRO financing | 0 | (0=%AP,1=LAUTRO |
| Financing Lautro | 0.0% | Lautro |
| Financing per AP | 196.7% | %AP |
| Finance Rate | 5.00% | 1.0041 |
| Repay | 12.46% | Mth Prem |

**Figure 1 Difficult input parameters**

*2.2 Auditability*

The answers produced by models depend on the inputs; if they cannot be checked it is impossible to say whether the answers are correct. Many models are implemented through a series of linked spreadsheets, and it is impossible to tell whether one of these spreadsheets is correct without being able to check the data imported from the others. The term "linked" is used in its broadest sense here, to cover any situation in which one spreadsheet uses results produced by another. Sometimes automatic links are used; often data is simply copied and pasted by hand, or, even worse, typed in from hard copy. In many cases it is impossible to check whether the correct numbers have been used.

As an example of how bad this can be, a spreadsheet that I reviewed recently used rates calculated in another spreadsheet. The standard procedure was to open the other spreadsheet, change the parameters so that they were consistent with the model that was being used, copy and paste the results into the spreadsheet under review, and close the other spreadsheet, often without saving it. There was thus no record at all of the assumptions that had been used to calculate the imported data, and it was impossible to determine whether they were consistent with the rest of the model.

**3 OTHER CRITERIA**

Even when a spreadsheet is technically correct, in that it uses the right formulae, it may still produce the wrong numbers. This is because a spreadsheet never exists in isolation; it is always part of a larger system, consisting of at least the spreadsheet and a user, and possibly other components as well. The interactions between the user, the spreadsheet, and the other components are a vital part of the operation of the system as a whole.

*3.1 Usability*

Probably the most common cause of wrong results from a technically correct spreadsheet is poor usability, which can affect the results in the following ways:

- Poor layout or misleading labels can make the user misread or misinterpret the results. In the end, it's not the results that appear in the spreadsheet that matter, but those that are used.

- In some spreadsheets it is nearly impossible for the user to specify the correct input parameters. For example, in the spreadsheet shown in Figure 1 you might think that in order to have NPP financing you should simply set the flag to zero. In fact, you also have to change a number of other parameters.

![Figure 1: Screenshot of a spreadsheet showing a complex formula]

**Figure 1 A complex formula**

- The users of this spreadsheet confirmed they did sometimes forget to change all the necessary parameters, and that this had led to problems in the past.

- Any spreadsheet that uses manual procedures in the course of calculating the results may produce errors when users forget to perform the procedures, or make mistakes as they do so. A common source of problems in this category is macros that are used to perform calculations. It is often very easy for the user to change the input parameters but not to run the macro: the spreadsheet is then in an inconsistent state. Another source of problems is pasting values into the spreadsheet. Recently Transalta Corporation reported a loss of $24m due to an error of this type [Cullen 2003]

- One of the classic problems with spreadsheets is that users simply overtype formulae with numbers without realising what they are doing.

*3.2 Maintainability*

Spreadsheets rarely stay the same: that's often one of the reasons for using the technology in the first place. Users may need to add functionality or to update the formulae that are used. It's important that they can do so quickly and easily without adding new errors. Unfortunately, many spreadsheets are built in such a way that even if they are correct to start with, they are unlikely to stay that way for long. Possible problems include:

- Hard coded constants. It is very common to see the same numbers, such as a maximum age of 65, used in many different formulae. Changing the age then means looking through every single formula to see if it contains any numbers that could derive from 65.

- Long or complex formulae, especially those containing many nested IF statements, are difficult to understand and change. The formula shown in Figure 1has seven conditional branches.

- Undocumented VBA code. One spreadsheet that I recently reviewed had 8000 lines of undocumented, unindented code. The chances of its being correct in the first place were slim; of staying correct after any changes, vanishingly small.

Interestingly enough, for spreadsheets, maintainability is closely related to usability (this is not the case in many programming paradigms). This complicates the situation as it is sometimes difficult to achieve good usability for all the different types of user at the same time. However, there are many techniques that make life easier for all users, including most of those recommended in [Raffensperger 2000].

*3.3 Performance*

Performance, like usability, is often considered to be primarily a productivity issue. Of course it does have significant productivity implications, but poor performance can also lead to erroneous results.

- Slow recalculation times or macro execution times can mean that even less testing is performed than would otherwise be the case. Rather depressingly, this rarely has any effect as no testing is performed anyway, despite the business critical nature of most spreadsheets.

- Slow calculation of either sort can also mean that the solution space is not explored effectively. Many results turn out to be extremely sensitive to small differences in input parameters, and it is often important to try many different sets of input parameters to see how stable the results are.

- Slow recalculation times can encourage the use of the manual recalculation setting in Excel. This is especially dangerous as it affects all open workbooks, not only the one that inspired its use, and it is very easy to forget to recalculate manually. The spreadsheet is then in an inconsistent state, as the results do not reflect the inputs.

It is surprising how often simple improvements can make large differences to performance. For example, recent reviews have resulted in improving macro execution from 359 seconds to 29 seconds simply by reading from and writing to arrays instead of one cell at a time, and recalculation from 10.2 seconds to 0.4 seconds by cutting down on the use of VLOOKUP.

## 4 PROBLEMS AND SOLUTIONS

Although spreadsheets are generally designed and implemented by people with no training in software engineering they are, of course, just another type of software system. The widespread use and business criticality of spreadsheets are not matched by a corresponding awareness of how the use of simple software engineering techniques can minimise the significant risks attached to their use [Pryor 2002b].

Typically, people tend to overestimate the costs of "doing things properly" and underestimate the benefits. For example, an actuary recently told me "In a 1% world, we can't afford to test our spreadsheets properly" (1% refers to the level of expenses for stakeholder pensions). I have also been told that documentation takes too long, and that it's impossible to keep track of different versions of the same spreadsheet.

In addition, many people are unaware of the many useful facilities that Excel provides. In one team I worked with, nobody had previously encountered either the data validation function or conditional formatting. Advanced techniques such as dynamic ranges or array formulae were quite out of the question. In many cases spreadsheet development is not seen as a skill that is valuable to the employer. Although it is common for people to spend as much as 80% or 90% of their time working with spreadsheets, development skills do not form part of their appraisals.

There are three steps to addressing the problems. First, people must know that the problems exist. Both spreadsheet developers and those who use the results they produce, including senior management, should be aware of the risks that spreadsheets pose, and that those risks need to be controlled. The achievement of this step is helped, in the financial services industry at least, by the emphasis that the Financial Services Authority places on the management of risk, including operational risk, and the importance of involving senior management.

Second, people must know what they can do to control the risk. Training is an important component of this step. Much of spreadsheet training is concerned solely with the techniques that are available, not with how and when they should be used. In the workshops that I lead, the emphasis is on the criteria for good spreadsheets, and how the techniques that we discuss can be used to achieve those criteria.

Third, people must actually take the required actions. This means having effective processes and standards that, far from being burdensome to the people that implement them, are perceived to provide both short and long term benefits. It also means having tools to support those standards and processes, such as XLSior, an Excel add-in that provides automated testing and documentation facilities (see http://www.xlsior.com).

## 5 CONCLUSIONS

It is well known that spreadsheets have very high error rates. However, it is often difficult to determine their correctness or otherwise as they are neither specified nor auditable. This paper argues that specification and auditability are two important criteria by which to judge spreadsheets, along with correctness, usability, maintainability and performance. Correctness alone is not enough to ensure the right results. Unsurprisingly, these criteria are among the factors generally accepted as indicating software quality [Pressman & Ince 2000].

Achieving the criteria requires knowledge of them and the risks associated with failing to achieve them, awareness of actions that can be taken to achieve them and control the risks, and the use of processes, standards and tools that support the performance of those actions.